\documentclass[aps,prl,twocolumn]{revtex4}
\usepackage{amsmath}
\usepackage{amssymb}
\usepackage[dvips]{graphicx}
\usepackage{wrapfig}
\usepackage{subfigure}
\usepackage{psfrag}

\newcommand{\sign}{\textrm{ sign}\,}

\begin{document}
\DeclareGraphicsExtensions{.eps}
\title{`Local' Quantum Criticality at the Nematic Quantum Phase Transition 
}
\author{Michael J. Lawler}
\author{Eduardo Fradkin}
\affiliation{Department of Physics, University of Illinois at Urbana-Champaign, 1110 W. Green Street, Urbana, Illinois 61801-3080, U.S.A.} 
\date{\today}

\begin{abstract}
We discuss the finite temperature properties of the fermion correlation function near the fixed point theory of the charge nematic quantum critical point (QCP) of a metallic Fermi system. We show that though the fixed point theory is above its upper critical dimension, the equal time fermion correlation function takes on a universal scaling form in the vicinity of the QCP.  We find that in the quantum critical regime, this equal-time correlation function has an ultra local behavior in space, while the low-frequency behavior of the equal-position auto correlation function is that of a Fermi liquid up to subdominant terms.  This behavior should also apply to other quantum phase transitions of metallic Fermi systems. 
\end{abstract}

\maketitle

A large number of metallic strongly correlated systems exhibit the phenomenon of quantum criticality. Typically the quantum critical regime is accessed either by continuously tuning an external control parameter (e.g. magnetic field or pressure), or by chemical means (e.g. doping). This phenomenon has been observed, among others, in heavy fermion systems, itinerant ferromagnets, ruthenates, in CDW systems, in high $T_c$ cuprate superconductors, and other complex oxides.

The theory of quantum phase transitions in metallic systems goes back to Hertz's seminal work\cite{Hertz76} and its subsequent development by Millis\cite{Millis93}. It yielded a  classification of these QCPs closely analogous to the theory of equilibrium classical critical phenomena. It is an order parameter theory which focuses on the behavior of its collective modes, with the added essential physical effects of gapless fermionic excitations in the quantum dynamics and as a mechanism for dissipation. A key assumption of this theory is that the fermions, although gapless, contribute primarily to dissipation mechanisms (e.g. Landau damping) and determine the effective dynamic critical exponent $z$. Following this logic one expects that at a QCP  the system should be scale invariant and the behavior of all its observables be controlled by some fixed point theory. Although this is indeed correct, the actual behavior of the fermionic quasiparticles (and their very existence) has remained a major unsolved problem. Naively one would have expected that the fermion correlators should obey scaling as well, perhaps with a non-trivial anomalous dimension, as in one-dimensional Luttinger liquids. We shall see that this is not the case for dimension $d>1$.

Experiments in quantum critical metallic systems show that these QCPs exhibit a phenomenon known as `local' quantum criticality\cite{Si01}.  There is strong experimental evidence for scaling behavior in both frequency and momentum, with a characteristic  dynamic scaling exponent $z$, of the correlation functions of the relevant order parameter and its associated susceptibilities, and of all other correlations in the bosonic sector of these systems. Thus, the spin correlation functions measured by neutron scattering and the current and density correlation functions measured by light scattering, do exhibit scaling. In contrast, at these same QCPs, the fermionic correlators  (measured in angle resolved photoemission (ARPES)) typically exhibit scaling in frequency but show essentially no scaling in momentum, thus the term `local quantum criticality'. The explanation of these seemingly disparate behaviors has been a long standing puzzle.

In this paper we develop a fully solvable fixed point theory of the of the quantum phase transition between a Fermi liquid (FL) and a charge nematic Fermi fluid (N)\cite{Oganesyan01}. 
A charge nematic Fermi fluid is a translationally invariant state of a metallic system which breaks spacial rotational invariance (or the point group symmetry of the lattice) spontaneously\cite{Kivelson98}. 
A charge nematic state has been seen in experiments in quantum Hall systems\cite{Lilly99,Du99,Cooper02} (a QCP has not yet been observed). There is also convincing evidence for a charge nematic phase in  the ruthenate Sr${}_3$Ru${}_2$O${}_7$ (including quantum phase transitions)\cite{Grigera04}. It has also been suggested\cite{Varma05} that the `hidden order' phase of the heavy fermion compound URu${}_2$Si${}_2$ may be due to a p-wave analog of the charge nematic state.\cite{spin-nematic} We will show here that the fixed point theory of the nematic QCP exhibits the hallmark of local quantum criticality: anisotropic scaling (with $z=3$) for the bosonic excitations, while the fermions show scaling in frequency and no scaling in momentum (up to the effects of irrelevant operators). In fact, at finite temperature, we find that the behavior of the equal-time fermion correlator is strikingly similar to the results found in Ref.\cite{Ghaemi05} for vortex-like operators in the (seemingly unrelated) quantum Lifshitz model\cite{Ardonne04}. The concept of local quantum criticality, has been the focus of recent work using perturbative methods\cite{Chubukov05a}. We will attack this problem using non-perturbative bosonization methods\cite{Haldane94,Houghton93,CastroNeto93,CastroNeto95,Lawler06}. Both approaches agree in the perturbative regime.

The (Pomeranchuk) transition to a charge nematic Fermi fluid is a quantum phase transition in which the shape of the Fermi surface (FS) acquires a spontaneous quadrupolar distortion, and breaks rotational invariance spontaneously down to $\pi$-rotations. It is the simplest example of an electronic liquid crystal phase\cite{Kivelson98}. It has dynamic critical exponent $z=3$ so that in two dimensions, the fixed point theory is above its upper critical dimension.  A simple computation to one-loop in the RPA interaction done in Ref.\cite{Oganesyan01} obtained a fermion self-energy ${\mathcal Im}\Sigma(q=q_F,\omega)\propto\omega^{2/3} \gg \omega$ at low energies. By a scaling analysis, the interactions are thus relevant by a power of $1/3$. Hence, an understanding of the fermionic properties of this QCP requires a non-perturbative treatment.  

We recently used\cite{Lawler06} the method of high dimensional bosonization\cite{Houghton93, CastroNeto93, CastroNeto94, CastroNeto95, Haldane94,Houghton00} to develop a non-perturbative fixed point theory of the nematic order parameter and of the fermion propagator. We demonstrated explicitly that this is a non-Fermi liquid with a fermion residue $Z$ which vanishes at the QCP, while the single particle density of states, $N(\omega)$, remains essentially unchanged from its FL value.  We also showed that the bosonization result for the fermion propagator reproduces the perturbative results of Refs.\cite{Oganesyan01,Nilsson05}, and showed that it treats all rainbow diagrams and vertex corrections on an equal footing. 

In this letter, we discuss its finite temperature behavior.  We show that in the quantum critical regime and at the level of the fixed point theory, the equal-time fermion correlations are ultra local in space due to an infrared divergence which wipes out all correlations in space to all distances.  This divergence is controlled by dangerous irrelevant operators which then set the correlation length of the fermions. In the nematic phase, symmetry protects the Goldstone modes at finite temperatures leading to the conclusion that the spacial correlation length of the fermions vanishes in the thermodynamic limit. Despite this extreme behavior, the correlations at equal-positions and at long times remain well behaved and the single particle density of states, $N(\omega)$, is essentially unaffected. We also show that the equal-times function obeys a scaling form and utilize this behavior to describe the collapse of the FL as a critical phenomenon. 

We take as our starting point, a low energy effective Landau model of two-dimensional spinless fermions with a quadrupole density interaction:
\begin{equation}\label{eq:Hnematic}
	{\mathcal H} = \int_{|{\bf q}|-k_F < \lambda} \frac{d^2q}{(2\pi)^2}\left[
		\xi_q \hat n_{\vec q} + f_2(q) \hat {\bf Q}({\vec q})\cdot\hat{\bf Q}(-{\vec q})\right],
\end{equation}
where $\hat n_{\vec q}=\hat c^\dagger_{\vec q} \hat c_{\vec q}$ is the fermion occupation number operator, $N(0)f_2(q)=F_2/(1+\kappa |F_2| q^2)$ is an interaction of range $\sqrt{\kappa}$, 
\begin{equation}
\hat{\bf Q}({\vec q}) = \frac{1}{k_F^2}\int \frac{d^2k}{(2\pi)^2}
\begin{pmatrix} k_x^2-k_y^2 & 2k_xk_y \\
2k_xk_y & k_y^2-k_x^2 \end{pmatrix}
\hat c^\dagger_{\vec k - \vec q}\hat c_{\vec k}
\end{equation}
is the Fourier transform of the nematic order parameter field, the quadrupolar density, and $\xi_q$ is the quasiparticle energy which is kept up to cubic order in $|\vec q|-k_F$. This model was proposed in Ref.\cite{Oganesyan01} to study a transition to a nematic Fermi fluid which in mean field occurs for $N(0)f_2(0) = F_2 \leq -1$.  The behavior that we discuss here should also apply to lattice systems at their QCP\cite{Kee03,Dellanna06}, but not in the ordered phase. 
Although models of this type do not account the effects of orbital and spin degrees of freedom, important to the physics of real metals, it captures the basic physics of a quantum phase transition in a metallic Fermi system. As we shall see below, it describes the phenomenon of `local quantum criticality' observed in many experiments.

The bosonization of dense Fermi systems, relativistic or not, in $d>1$ space dimensions requires the introduction of local coordinate frames near the FS to describe the fluctuations of this smooth curved quantum object. Much as in the conventional Landau theory of the Fermi liquid\cite{Baym91}, the fluctuations of the FS describe the quantum diffusion of particle-hole wave packets on the FS\cite{CastroNeto93}. We therefore introduce a set of $N$ patches of width $\Lambda = 2\pi k_F/N \sim \sqrt{k_F\lambda}$ given by the curvature of the surface and cutoff with $N\to\infty$ in the renormalization group (RG) limit of $\lambda\to0$. Thus, all corrections of order $1/N$ are uninteresting within this framework provided they correspond to irrelevant operators. A key feature of the bosonization treatment is that the corrections to the linear dispersion (locally normal to the FS) are described by non-linear operators (in the bosons) which are irrelevant in the Landau phase and at the fixed point theory. Processes which involve large(er) momentum transfers tangent to the FS are described by both intra-patch and inter-patch interactions. Contrary to some claims\cite{Kopietz97,Chubukov06}, bosonization treats the effects of the curvature of the FS  exactly. 

Introducing local coordinates for patch $S$, located at Fermi wavevector $\vec k_S$, we find that this phase space analysis allows us to expand the fermion operator, $\hat \psi(\vec x)$, in terms of a set of fermion species derived from each patch:
\begin{equation}
	\hat\psi(\vec x) = \sum_{S} \hat\psi_S(\vec x)\frac{e^{i\vec k_S\cdot\vec x}}{\sqrt{N}},\    
\hat\psi_S(\vec x) = \int_{\vec q\in{\mathcal P}_S}\frac{d^2q}{(2\pi)^2}\ \hat c_q\, e^{i\vec q\cdot\vec x}
\end{equation}
In terms of the local Fermi velocity $\vec v_S$ at patch $S$, a vector locally normal to the FS, the dispersion is 
\begin{equation}
		\xi_q = \vec v_S\cdot\vec q + \beta(\vec v_S\cdot\vec q)^2 + \gamma(\vec v_S\cdot\vec q)^3.
		\label{eq:dispersion}
\end{equation}
where $|\vec v_S|=v_F$ and $|\vec q| \ll k_F$.  Due to this dispersion, our fluctuating FS 
can be thought of as $N\times\infty$ fermion species each with its own conserved charge due to the $U(1)^{N\times\infty}$ symmetry of the Landau theory and of Eq.\eqref{eq:Hnematic}:
\begin{equation}\label{eq:U1infty}
		\hat\psi_S(x_n,x_t) \to e^{i\theta_S(x_t)}\hat\psi_S(x_n,x_t)
\end{equation} 
where $(x_n,x_t)$ are local orthogonal patch coordinates with $x_n$ parallel to $\vec v_S$.  Though we have presented this within the language of our local coordinates, this charge conservation for each point of the FS is a well known property of the Landau Hamiltonian, {\it e.g.\/} Eq.\eqref{eq:Hnematic}, quite independent of the choice of coordinates on the FS.

A basic consequence of Eq.\eqref{eq:U1infty} is a simplification of the fermion correlation functions. If one creates a fermion of type $(S,x_t'=0)$ at $(x_n'=0,t'=0)$, one must destroy another fermion of the same type at $(x_n,t)$ to have any overlap with the ground state. Hence, only one species may contribute to the single particle fermion Green function. After defining a windowing function $W_\Lambda(x_t)$ that vanishes for $|x_t|\Lambda\gtrsim1$ and attains the value of unity at $x_t=0$, the correlator becomes
\begin{multline}
	G_F(\vec x,\tau) =\\ \sum_S\frac{W_\Lambda(x_t)}{N} e^{i k_F x_n}
      \langle G | T_{\tau} \psi_S(x_n,0,\tau)\psi^\dagger_S(0,0,0)|G\rangle
\end{multline}
where $x_n=\hat k_S \cdot \vec x$ and $x_t\perp x_n$. Following standard bosonization techniques we compute $G_F(\vec x,\tau)$ for $T>0$:
\begin{equation}
\label{eq:GF}
	G_F(\vec x,\tau) = \sum_S \frac{W_\Lambda(x_t)e^{ik_f x_n}}{N}G_{F(S)}^0(x_n,\tau){\mathcal Z}_S(x_n,\tau)
\end{equation}
where $G_{F(S)}^0(x_n,\tau)$ is the `patch' free Green function and 
\begin{multline}\label{eq:Zs}
  \ln{\mathcal Z}_S(x_n,\tau) =\\ -\int_{{\mathcal P}_S}\!\frac{d^2q}{(2\pi)^2}\sum_{n\in{\mathbb Z}}\frac{1}{\beta}
    \left(e^{i(k_nx_n-\omega_n\tau)}\!-\!1\right)
			\frac{V^{RPA}_{S,S}(\vec q,i\omega_n)}{\left(i\omega_n\!-\!\vec v_S\cdot\vec q\right)^2}
\end{multline}
with $V^{RPA}_{S,S}$ the effective RPA interaction.

The limit $N \to \infty$ can be obtained. Consider the case when ${\mathcal Z}_S$ is independent of the patch index $S$ (true for a circular FS) and decays or oscillates slower than $e^{ik_Fx_n}$ as $x_n\to\infty$. Converting the sum over patches to an integral, and using steepest descents, we find that since the integral is peaked around $\theta\equiv\cos^{-1}(\vec k_S\cdot\vec x/k_Fr)=0$, we get the angular average $\langle e^{ik_Fr\cos\theta}\rangle = J_0(k_F r)$, and 
\begin{equation}
  G_F(r,t) \sim G_F^0(r,t){\mathcal Z}_{S=0}(r,t)
\end{equation}
with $G_F^0(r,t)\sim \cos(k_Fr)/r^{3/2}$; the weighting function $W_\Lambda(x_t)$ and sum over patches have dropped out. 

Near the nematic QCP of \eqref{eq:Hnematic}, $V^{RPA}_{S,S}$ is directly related to the propagator of the order parameter field.  Although for angular momentum channels greater than s-wave, the order parameter theory contains a variety of collective modes, it has been shown\cite{Lawler06} that $\ln{\mathcal Z}_S$ is always dominated by the softest of these modes, a $z=3$ over damped mode very similar to that occurring in Hertz's original treatment of the ferromagnetic problem. 

Following Refs.\cite{Hertz76,Millis93,Oganesyan01}, and using standard results of the well known nematic-isotropic classical phase transition\cite{DeGennes93}, we construct the qualitative phase diagram for the Nematic quantum phase transition of a Fermi fluid.
There are clearly three regimes of interest. Two of them are disordered and described by
\begin{equation}\label{eq:VrpaSym}
	V^{RPA}_{S,S}(\vec q,i\omega_n) = \frac{1/N(0)}{\frac{|\omega_n|}{qv_F}-\kappa q^2 -\Delta}
\end{equation}
where $\Delta=\Delta(\delta,T)$ with $\delta = 1+1/F_2$, the distance to the mean field position of nematic QCP at $F_2=-1$. In the FL regime, $\Delta$ is dominated by the scale $\delta$ and we obtain $\Delta\approx\delta - \delta_C$ where $\delta_C$ arises from finite corrections to the gaussian theory due to irrelevant interactions. In the quantum critical regime, $\Delta$ is dominated by temperature and we must resort to a careful renormalization group argument to understand its behavior\cite{Millis93}. 
We find that the temperature dependence of $\Delta$ follows the law
 $ \Delta = -\gamma T\ln (T)$,
where $\gamma$ is the coupling to the dangerous irrelevant quartic interactions of the order parameter theory, which in the model of Eq.\eqref{eq:Hnematic} is the cubic term in the dispersion relation,\cite{Oganesyan01,Barci03} Eq.\eqref{eq:dispersion}.
In the nematic phase, the Goldstone mode dictates that $\Delta=0$, 
\begin{equation}\label{eq:VrpaNem}
	V^{RPA}_{S,S}(\vec q,i\omega_n) = 
    \frac{1/N(0)}{\frac{|\omega_n|}{qv_F}-\frac{\kappa}{\sin^22\theta_S} q^2}
\end{equation}
In contrast to Eq.\eqref{eq:VrpaSym}, it depends on the patch index $S$.

We now
compute $G_F(\vec x,\tau)$ as a function of $T$ and $\delta$, near the QCP. 
We are interested in the behavior of the amplitude ${\mathcal Z}_S$, discussed at $T=0$ in Ref.\cite{Lawler06}. 
For a FL,
\begin{equation}
  Z_S(x_n,t=0)|_{x_n\to\infty} \equiv Z < 1,\quad
  Z_S(0,t)|_{t\to\infty}=\textrm{const}
\end{equation}
At the QCP, the leading behavior is
\begin{equation}
   Z_S(x_n,t=0)|_{x_n\to\infty}\sim e^{-(x_n/R)^{1/3}},\;
   Z_S(0,t)|_{t\to\infty}=\textrm{const}
\end{equation}
There is a similar non Fermi liquid behavior in the nematic phase, except that there are four special points at $\sin2\theta_S=0$ where the leading behavior is that of a FL. 

For $T>0$, 
the $\omega_n=0$ contribution to ${\mathcal Z}_S$ in Eq.\eqref{eq:Zs} dominates for $T_\kappa \ll T \ll T_F$, $T_\kappa = v_F/k_B\kappa$, and
\begin{align}\label{eq:highT}
	\ln{\mathcal Z}_S\big|_\textrm{{static}} &= \frac{k_BT}{N(0)\kappa}\int_{{\mathcal P}_S}\frac{d^2q}{2\pi^2}
			\frac{\cos(q_nx_n) - 1}{v_F^2q_n^2(q^2+\Delta/\kappa)}\\
		&\approx \left\{\begin{array}{ll}
      -\frac{k_BT}{N(0)v_F^2\kappa}x_n^2\log(\xi/x_n) & 0\leq x_n\ll\xi\\
      -\frac{k_BT}{N(0)v_F^2\kappa}|x_n|\xi             & x_n\gg\xi
    \end{array}\right.
\end{align}  
where $\xi=\sqrt{\kappa/\Delta}$. On the FL side, $\xi$ is quite small and we obtain exponentially decaying spacial correlations. In the quantum critical regime, $\xi$ is very large, set only by the irrelevant interactions and temperature.
Finally, in the nematic phase  $\xi=\infty$ for $T>0$, and 
${\mathcal Z}_S\to0$, kept finite only in a finite sized system (except once again at the four special points).  At equal positions ($x_n=0$), the $\omega_n=0$ contribution vanishes, and the equal position auto-correlation function is well behaved in the infrared.

Near the QCP, $\ln{\mathcal Z}_S(x_n,0)$ obeys the scaling law:
\begin{equation}\label{eq:scaling}
	\ln{\mathcal Z}_S(x_n,0, \delta, T) = b^{-1/3}\ln{\mathcal Z}_S(b x_n,0, \delta/b^{2/3}, T/b)
\end{equation}
where 
all length scales  
are measured in units of the interaction range $\sqrt{\kappa}$. Upon the change of variables
 $ q_n' = q_n/b,\quad q_t'=q_t/b^{1/3}$,
as $b\to0$, $q\to q'b^{1/3}$, 
\begin{equation}
	V^{RPA}_{S,S}(q_n'b,q_t'b^{1/3},i\omega_n'/b)\to b^{-2/3}V^{RPA}_{S,S}(q_n=0,q_t',i\omega_n')
\end{equation}
For $T=\delta=0$ we find,
\begin{equation}
	\ln{\mathcal Z}_S(x_n,0, 0, 0) = |x_n|^{1/3}\ln{\mathcal Z}_S(\sign(x_n),0, 0,0)
	\label{eq:Z0}
\end{equation} 
by letting $b=|x_n|^{-1}$. Since $\ln{\mathcal Z}_S(\pm1,0,0,0)<0$, we find ${\mathcal Z}_S(x_n,0,0,0)\sim e^{-a|x|^{1/3}}$ as expected. From
\begin{equation}\label{eq:residue}
	Z = \lim_{x_n\to\infty} {\mathcal Z}_S(x_n, 0 ,\delta,T) = 
    \lim_{q\to0} \left(n_{k_F-q}-n_{k_F+q}\right)
\end{equation}
and Eq.\eqref{eq:Z0}, we find $Z=0$ at the QCP. As $\delta\to 0$ at $T=0$, by letting $b=\delta^{3/2}$ we get
\begin{equation}
 Z  = \exp\left(\ln{\mathcal Z}_S(\infty,0,1,0)/\sqrt{\delta}\right)
\end{equation}
Thus, since $\ln{\mathcal Z}_S(\infty,0,1,0)<0$, $Z$ vanishes with an essential singularity as $\delta \to 0$. 
Thus, this scaling form, ${\mathcal Z}_S(x_n,0)$ clearly exhibits the breakdown of the FL as the QCP is approached. 

Let us complete our discussion of scaling by going to $T>0$. Letting $\Delta = \xi^{-2}$ and $b=T$ in \eqref{eq:scaling}, we obtain:
\begin{equation}
	\ln{\mathcal Z}_S(x_n,0,\xi^{-2},T) = T^{-1/3}\ln {\mathcal Z}_S\left(x_nT,0, \xi^{-2}T^{-2/3},1\right)
\end{equation}
Since we only need ${\mathcal Z}_S$ for $T>0$, we may evaluate it again using the $\omega_n=0$ mode (now in the scaling limit):
\begin{equation}
  \ln{\mathcal Z}_S(x_n,0,\xi^{-2},T) = -A |x_nT| \xi/k_F
\end{equation}
so that holding $x_nT$ fixed, this diverges like $(-T\ln T)^{-1/2}$ as $T\to0$ \emph{and therefore it is not a scaling function of $x_nT$}. Since the free fermion Green function in \eqref{eq:GF} asymptotically behaves as $e^{-\pi |x_nT|}$, we see that the above form dominates over the free fermion value for
 $ \frac{\xi}{k_F} > \pi$,
or for
$  T \lesssim T_g \equiv T_\kappa/ (gk_F^2\kappa) $,
where we have reintroduced $\kappa$ with $T_\kappa = v_F/\sqrt{\kappa}k_B$. Hence, for $T\gg T_g$, the free fermion result dominates and scaling occurs in $x_nT$. For $T\ll T_g$, scaling in $x_nT$ breaks down.

In this letter, we demonstrated non-perturbatively the phenomenon of local quantum criticality near the metallic charge nematic quantum phase transition. It appears to be a robust behavior of many quantum critical systems, as, in addition to the nematic quantum critical point, it also applies to the critical behavior of quantum dimer models\cite{Ghaemi05}, metalic antiferromagnets and ferromagnets\cite{Chubukov05a}. While the order parameter correlations, $\xi$ are large in the quantum critical regime, kept finite only by the dangerous irrelevant quartic interactions,
the fermionic correlations in this regime are correspondingly short ranged, kept non-zero by the same dangerous irrelevant quartic interactions.  Indeed, in the nematic phase, which has an infinite correlation length, we have shown that the corresponding fermionic correlations vanish in the thermodynamic limit except at four special points on the FS. An extension of this analysis to other quantum phase transitions in metallic systems results in  
the same basic conclusion, that the `local quantum criticality' of the fermions is independent of $z$ and it is due to the $\omega_n=0$ static contribution to the correlation function. 
We found that while the single particle density of states, $N(\omega)$ remains finite throughout the entire phase diagram,
the qp residue $Z$ (the jump in the fermion occupation number) 
vanishes for $T< T_g$ both in the quantum critical regime and in the nematic phase, which suggests that a FS may not be seen in a de Haas-van Alphen experiment. 

\noindent
{\bf Acknowledgments}:
We thank D. Barci, A. H. Castro Neto, A. Chubukov, V. Fern\'andez, S. Kivelson, V. Oganesyan, C. P{\'e}pin, and C. Varma for discussions. This work was supported in part by the NSF grant DMR 0442537, by an A. O. Beckmann Award of the Research Board of the University of Illinois, and by the DOE Award No. DEFG02-91ER45439 at the Materials Research Laboratory of the University of Illinois. 


\begin{thebibliography}{30}
\bibitem[{spi()}]{spin-nematic}
\bibinfo{howpublished}{The charge nematic state should not be confused with the
  spin nematic state discussed in Ref.\cite{gorkov92}.}
\end{thebibliography}

\end{document}